\documentclass[aps,prl,twocolumn,groupedaddress]{revtex4}
\usepackage{graphicx}

\begin{document}


\title{SU(4) versus SU(2) Kondo effect in double quantum dot}


\author{A. L. Chudnovskiy}
\email[]{achudnov@physik.uni-hamburg.de}
\affiliation{1. Institut f\"ur Theoretische Physik, Universit\"at Hamburg,
Jungiusstr. 9, 20355 Hamburg, Germany}


\date{\today}

\begin{abstract}
We consider the spin and orbital Kondo effect in a
parallel arrangement of two strongly electrostatically coupled quantum dots.
Increasing the exchange of electrons between the dots through the attached leads induces a transition
between the $SU(4)$ spin- and orbital Kondo effect and $SU(2)$ spin Kondo effect. Being the same for the
$SU(4)$ and $SU(2)$ symmetry points, the Kondo temperature drops slightly in the intermediate regime.
Experimentally, two kinds of Kondo effects can be discriminated by the sensitivity to the suppression of the spin Kondo effect by Zeeman field. The dependence of the Kondo temperature on Zeeman field and the strength
of electronic exchange mediated by the leads is analyzed in detail.
\end{abstract}


\maketitle


Kondo effect in quantum dot devices remains a subject of active theoretical and experimental
investigations since its first observation \cite{DGG}.
The observation of the Kondo effect with orbitally degenerate levels provided the first demonstration of the strong influence of the orbital structure of the states in the dot and attached leads on the Kondo effect \cite{Kouwenhoven2}.
Further exploration of the interplay between spin and orbital degrees of freedom in the Kondo effect became possible in experiments with double dot systems.
If the two dots are strongly electrostatically coupled \cite{Holleitner,Weis},  then
there are regions in the charging diagram of the double dot device, where there is no energy cost to transfer an electron between the two dots. In that regions, the two ground states
of the double dot system with occupations of the two dots $(N_1-1, N_2+1)$ and $(N_1, N_2)$
are degenerate. The two ground states build a two-dimensional Hilbert space which spans the
representation of the $SU(2)$ group, hence a spin-like degree of freedom called pseudospin
can be assigned to those two states. Quite analogously to
and to a large extent independently of the well-known spin Kondo effect,  the orbital fluctuations in transport through the double dot result in the development of the orbital, or pseudospin Kondo effect.
Furthermore, at special values of Kondo couplings, the combined spin and pseudospin Kondo Hamiltonian
possesses a $SU(4)$ symmetry with respect to rotations in the spin-pseudospin space. In that regime, the $SU(4)$  Kondo effect with greatly enhanced Kondo temperature has been predicted theoretically \cite{Borda,Zarand}.

The existence of the pseudospin Kondo effect crucially relies on the coupling of each quantum dot to its
own electronic reservoir formed by the electron states in the attached leads. The separation of the reservoirs allows
to define a pseudospin for the electrons in the leads in a natural way. For two sequentially coupled quantum dots \cite{Borda} such a separation is given by geometry. In contrast, the realistic experimental geometry for two dots coupled {\it in parallel} to the leads does not introduce a separate reservoir for each dot  {\it a priori} \cite{Holleitner}.

\begin{figure}
\includegraphics[width=8cm,height=6cm,angle=0]{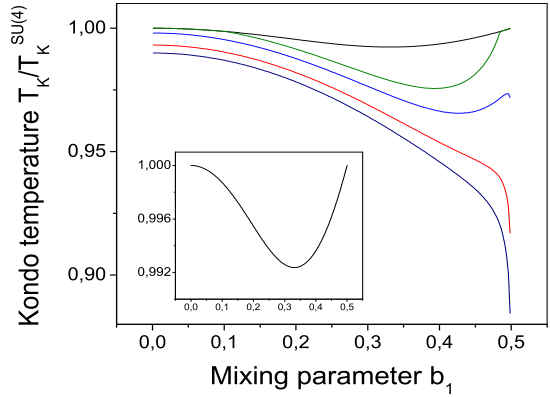}
\caption{Kondo temperature as a function of mixing $b_1$ for different Zeeman fields $h$. Magnetic field increases from the highest to the lowest curve. From the highest to the lowest curve: $h/T_K^{SU(4)}=1.0; 1.02; 1.06; 1.14; 1.29$.  Inset: the dependence $T_k(b_1)$ at zero Zeeman field. The curves illustrate characteristic $T_K(b_1)$ dependencies.
\label{figTk-b1}}
\end{figure}

In this letter, we investigate the possibility and properties of spin and pseudospin Kondo effect in a double dot embedded in a parallel circuit with the attached leads if there is no separate electron reservoirs for each quantum dot.
The separation between the two electronic reservoirs is characterized by the asymmetry of amplitudes of each electronic mode to tunnel from the reservoir into one or into the other dot.  We introduce a measure of mixing between the two reservoirs that reflects the structure of tunnel matrix elements. The mixing between the reservoirs can be varied either by changing the potential barrier that separates them or by applying the perpendicular magnetic field \cite{Lopez1}. Without the perpendicular magnetic field, in the limit of completely symmetric tunneling from each mode to both quantum dots, the Kondo effect in the orbital sector is suppressed. At the same time, all modes of both reservoirs couple to a single electron state given by a symmetric combination of the states localized in each dot. This results in the doubling of the density of states in the reservoir for the spin Kondo effect. At the end, the Kondo temperatures for the two symmetry points, the $SU(4)$ one and the $SU(2)$ one, are equal \cite{Zarand,Choi}. We find however that  the Kondo temperature gets slightly suppressed in the intermediate regime between the two symmetry points (see inset in Fig. \ref{figTk-b1}) independently of the method used to change the mixing between the two reservoirs (potential barrier or perpendicular magnetic field). This result should be contrasted with the analysis given in Ref. \cite{Lopez1}.   The change in symmetry of the Kondo effect can be detected experimentally by applying an external Zeeman magnetic field in the plane of the double dot device.  An external Zeeman field suppresses the Kondo effect in the spin sector. Therefore, whereas the $SU(2)$ Kondo effect is completely suppressed by the Zeemann field, the combined spin-pseudospin
$SU(4)$ Kondo effect is only reduced to the $SU(2)$ Kondo effect in the pseudospin sector. In what follows we derive an expression for the Kondo temperature as a function of mixing between the reservoirs and of a Zeeman magnetic field in the whole range between the $SU(4)$ and $SU(2)$ regimes.

We consider a device consisting of two single-level quantum dots coupled in parallel to
external Fermi liquid leads. The Hamiltonian represents a sum of the following terms: the Hamiltonian of isolated dots including the interdot interaction, the Hamiltonian of Fermi liquid reservoirs,
and the tunneling between the quantum dots and the source ("$s$") and drain ("$d$") leads
\begin{equation}
H=H_{QD}+\sum_{r=s,d}H^{res}_{r}+H^t.
\label{H1}
\end{equation}
The Hamiltonian of isolated quantum dots reads
\begin{equation}
H_{QD}=\sum_{i=1,2}\left[\sum_{\sigma=\uparrow\downarrow}E_i\hat{n}_{i\sigma}
+U\hat{n}_{i\uparrow}\hat{n}_{i\downarrow}\right]+U_{12}\hat{n}_1\hat{n}_2,
\label{Hqd}
\end{equation}
where the following notations are introduced: $\hat{n}_{i\sigma}=
\hat{c}^{\dagger}_{i\sigma}\hat{c}_{i\sigma}$ -- the number of electrons in the dot
$i$ with $z$-projection of spin $\sigma$ ($\sigma=\uparrow \mbox{or} \downarrow$),
$\hat{c}_{i\sigma}$ is an annihilation operator of
an electron in dot $i$ with spin projection $\sigma$. We focus on the $SU(4)$ Kondo
regime that  is achieved at $E_1=E_2=E$, $U_{12}=U$.
The low energy sector of the model consists of states with a total of one electron in the
double dot. Their energy in the isolated dots equals $E(<0)$.
The corresponding ket-vectors are denoted as $|\sigma,0\rangle$ and
$|0,\sigma\rangle$, where $0$ denotes the unoccupied dot. Single particle tunneling transforms the low energy states into the empty state $|0,0\rangle$ (energy $0$) or into states with a total of two electrons (energy $E+U (>0)$), which are $|\uparrow\downarrow,0\rangle$,
$|0,\uparrow\downarrow\rangle$, $|\sigma_1,\sigma_2\rangle$.
The excited states with a total of more than two electrons in a single dot do not couple
to the low-energy sector in second order in tunneling. They are omitted from consideration.

The tunneling Hamiltonian is given by
\begin{equation}
H^t=\sum_{i=1,2}\sum_{r=s,d}\sum_{k\sigma}
T_{ik}^{r}\hat{a}^{\dagger}_{r k\sigma}\hat{c}_{i\sigma}+h.c.
\label{tunneling}
\end{equation}
In the presence of perpendicular magnetic field, the tunnel matrix elements are complex. We choose a symmetric gauge: $T_{1k}^s=e^{i\phi/4}|T_{1k}^s|$, $T_{2k}^s=e^{-i\phi/4}|T_{2k}^s|$,
$T_{1k}^d=e^{-i\phi/4}|T_{1k}^d|$, $T_{2k}^d=e^{i\phi/4}|T_{2k}^d|$, where $\phi$ is the Aharonov-Bohm phase in presence of perpendicular magnetic field. To elucidate the appearance of the pseudospin in a general case, when each electronic mode has  nonvanishing tunneling amplitudes to both quantum dots, let us define the tunneling angle $\eta_k^{r}$ for each mode in the given reservoir by the following relations:
$
\cos\left(\eta_k^{r}\right)=\frac{\left|T^{r}_{1k}\right|}{T^{r}_k}$,
$
\sin\left(\eta_k^r\right)=\frac{\left|T^{r}_{2k}\right|}{T^{r}_k}$, where $T^{r}_k=\sqrt{\left|T^{r}_{1k}\right|^2
+\left|T^{r}_{2k}\right|^2}$.
Furthermore, we introduce formally bi-spinor notations for the operators in the fermionic reservoir and
in the double dot
\begin{eqnarray}
&&
\hat{\Psi}_{r k}=\left(e^{\mp i{\phi}/4}\cos\eta^{r}_k,e^{\pm i{\phi}/4}
\sin\eta^{r}_k\right)^T\otimes
\left(\hat{a}_{r k\uparrow},\hat{a}_{r k\downarrow}\right)^T , \label{Psi}\\
&&
\hat{\Phi}=\left(\hat{c}_{1\uparrow},\hat{c}_{1\downarrow},
\hat{c}_{2\uparrow},\hat{c}_{2\downarrow}\right)^T. \label{Phi}
\end{eqnarray}
In (\ref{Psi}) the upper signs in the exponent relate to $r=s$ and the lower ones to $r=d$.
Using the notations above,  we rewrite the tunneling Hamiltonian in the form
\begin{equation}
 H^t=\sum_{k}\left(\frac{|T_k^s|}{T_k}\hat{\Psi}_{sk}^{\dagger}+
 \frac{|T_k^d|}{T_k}\hat{\Psi}_{dk}^{\dagger}\right)\hat{\Phi}+ h.c.,
\label{HT}
\end{equation}
where $T_k=\sqrt{|T_k^s|^2+|T_k^d|^2}$.
One can see from (\ref{HT}) that for each $k$ it is only the mode $\Psi_k=\frac{|T_k^s|}{T_k}\hat{\Psi}_{sk}
+ \frac{|T_k^d|}{T_k}\hat{\Psi}_{dk}$ \cite{Glazman-Raikh}.
Performing the Schriefer-Wolff transformation for Hamiltonian (\ref{H1}), we derive the effective Kondo Hamiltonian describing the dynamics of the low-energy sector of the model that can be written as
\begin{equation}
H_K=\sum_{\mu,\nu=0}^3\sum_{k,k'} J^{kk'}_{\mu\nu}\hat{\Phi}^{\dagger}\left(\hat{\tau}^{\mu}\otimes\hat{\tau}^{\nu}\right)
\hat{\Phi}\hat{\Psi}_{k}^{\dagger}\left(\hat{\sigma}^{\mu}\otimes\hat{\sigma}^{\nu}\right)
\hat{\Psi}_{k'}+H_{res}.
\label{HK}
\end{equation}
Here $\hat{\Phi}^{\dagger}\left(\hat{\tau}^{\mu}\otimes\hat{\sigma}^{\nu}\right)
\hat{\Phi}$ and $\hat{\Psi}_{k}^{\dagger}\left(\hat{\tau}^{\mu}\otimes\hat{\sigma}^{\nu}\right)
\hat{\Psi}_{k}$ represent the ``hyperspins'' in the pseudospin-spin $SU(4)$ space for the double dot and
fermionic reservoir respectively, $\hat{\tau}^{\mu}$ and $\hat{\sigma}^{\mu}$ denote the corresponding Pauli matrices. For the $SU(4)$ symmetry point, the Kondo couplings equal $J_{\mu\nu}^{kk'}=T_kT_{k'}\left(\frac{1}{E+U}-\frac{1}{E}\right)$ for all $\mu,\nu \in\{0,... ,3\}$ except $\mu=\nu=0$. In what follows we assume the $|T^r_k|$ to be independent of $k$, hence we omit the indeces $k,k'$ at $J_{\mu\nu}$.
Note the principal difference between the hyperspins in the quantum
dot and in the reservoir. Whereas operator $\hat{\Phi}$ incorporates four dynamical degrees of freedom,
the operator $\hat{\Psi}_{k}$ does only two, $\hat{a}_{rk\uparrow}$ and $\hat{a}_{rk\downarrow}$. The
angle $\eta^r_k$ is not dynamical. Yet this angle can fluctuate with wave vector $k$.
The magnitude of those fluctuations determines whether the pseudospin in reservoirs is promoted to a
dynamical degree of freedom. For example, if the reservoirs for two quantum dots are strictly separated for both source and drain leads, then the angle $\eta^r_k$ assumes two possible values $0$ and $\pi/2$, which results in strong fluctuations
of $\eta^r_k$ with $k$, and eventually leads to the pseudospin Kondo effect. In contrast, if each lead, the source and the drain, contains only a single common reservoir  for both dots, then, without the perpendicular magnetic field ($\phi=0$)  the tunneling amplitudes to both dots are equal
$T^r_{1k}=T^r_{2k} \ \forall k$. In that case, the angle $\eta^r_k$ is frozen at the value $\pi/4$ for any $k$ and $r$, and the orbital Kondo effect is suppressed.

\begin{figure}
\includegraphics[width=4cm,height=1.5cm,angle=0]{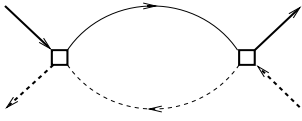}
\caption{One loop renormalization of Kondo coupling.  Solid line: the propagator of
electrons in the reservoir $\hat{G}_r(\omega,k)$. Dashed line: the spin-propagator in the double
dot $\hat{D}(\omega)$, see (\ref{D}), (\ref{G}).
 \label{fig-1loop}}
\end{figure}

In what follows we present results of the renormalization group (or poor man scaling)  analysis of  the Hamiltonian (\ref{HK}). We consider the case of equal coupling strengths to the source and drain, that is $|T^s_{ik}|=T^d_{ik}|$, hence $\eta_k^s=\eta_k^d$. The one loop renormalization of the Kondo couplings $J_{\mu\nu}$ is given by the diagram in Fig. \ref{fig-1loop} and the diagrams obtained by change of direction of arrows for solid or dashed lines.
The propagators of the field in the double dot $\hat{D}(i\omega_n)$ and
in the electronic reservoir $\hat{G}(i\omega_n,k)$ are given by
\begin{eqnarray}
&&
\hat{D}(i\omega_n)=1/\left(i\omega_n\right) \left(\hat{1}_2\otimes\hat{1}_2\right), \label{D} \\
&&
\nonumber
\hat{G}(i\omega_n,k)=\frac{1}{i\omega_n-\xi_k}\times \\
&&
\left[
\begin{array}{cc}
\cos^2\eta_k & \cos(\phi/2)\cos\eta_k\sin\eta_k \\
\cos(\phi/2)\cos\eta_k\sin\eta_k & \sin^2\eta_k
\end{array}
\right]\otimes \hat{1}_2, \label{G}
\end{eqnarray}
where $\eta_k^s=\eta_k^d=\eta_k$.
One can see that in comparison to the pure $SU(2)\otimes SU(2)$ Kondo effect as predicted for sequentially coupled dots, the presence of mixing between the reservoirs results in a nontrivial structure of the fermion propagator $\hat{G}(i\omega_n,k)$ in orbital space.
After integration in the infinitesimal energy shell between $\Lambda-\delta\Lambda$ and $\Lambda$, $\Lambda$ being the  high energy cutoff,  the correction to the Hamiltonian can be written as
\begin{widetext}
\begin{eqnarray}
\nonumber &&
\delta \hat{H}_K
=\pi\nu_F \left(\delta\Lambda/\Lambda\right) J_{\mu\nu}J_{\mu'\nu'}\hat{\Phi}^{\dagger}
\left[\left(\hat{\tau}^{\mu}
\otimes\hat{\tau}^{\nu}\right)\left(\hat{\tau}^{\mu'}\otimes\hat{\tau}^{\nu'}\right)-
\left(\hat{\tau}^{\mu'}\otimes\hat{\tau}^{\nu'}\right)\left(\hat{\tau}^{\mu}
\otimes\hat{\tau}^{\nu}\right)\right]\hat{\Phi}\times \\
&&
\hat{\Psi}^{\dagger}\left[\left(\hat{\sigma}^{\mu}\otimes\hat{\sigma}^{\nu}\right)\hat{P}
\left(\hat{\sigma}^{\mu'}\otimes\hat{\sigma}^{\nu'}\right)-
\left(\hat{\sigma}^{\mu'}\otimes\hat{\sigma}^{\nu'}\right)\hat{P}
\left(\hat{\sigma}^{\mu}\otimes\hat{\sigma}^{\nu}\right)
\right]\hat{\Psi}.
\label{deltaHK}
\end{eqnarray}
\end{widetext}
Here $\nu_F$ denotes the density of states at the Fermi level,
$
\hat{P}=\left(b_0\hat{1}_2+\sum_{p=1}^3b_p\hat{\sigma}_p\right)\otimes \hat{1}_2
\label{P}
$, $b_0=1/2$, and the values of other constants $b_p$ are given by the following averages over the wave vector  $k$
\begin{eqnarray}
&&
b_1=\cos(\phi/2)\langle\cos\eta_k\sin\eta_k\rangle, \label{b1}\\
&&
b_2=0, \ \mbox{for $\eta_k^s=eta_k^d$}, \label{b2} \\
&&
b_3=\frac{1}{2}\langle\cos^2\eta_k-\sin^2\eta_k\rangle. \label{b3}
\end{eqnarray}
Due to the nontrivial structure of the fermion propagator new interactions are generated in course of the renormalization group (RG) transformation. The new interactions have the structure $J_{\mu\lambda\nu}\bar{\Phi}\left(\tau^{\mu}\otimes\tau^{\nu}\right)\Phi\cdot
\bar{\Psi}\left(\sigma^{\lambda}\otimes\sigma^{\nu}\right)\Psi$, where all the indices $\mu$, $\lambda$
and $\nu$ change independently.

We consider the case where the tunnel couplings to the two dots are symmetric, that is $b_3=0$. The mixing between the two reservoirs is now determined by the value of the parameter $b_1$. $b_1=0$ corresponds to the case of two strictly separated reservoirs, where the $SU(4)$ Kondo effect is expected, whereas the maximal value $b_1=b_0=1/2$ corresponds to a single common reservoir for both quantum dots. In the latter case only the spin $SU(2)$ Kondo effect is possible. The chosen form of the operator $\hat{P}$ with $b_1\neq 0$ distinguishes the coupling constants with $\mu,\lambda=\overline{0,1}$. It turns out that the only coupling constants with $\mu\neq\lambda$ that are generated by the RG-transformation are $J_{10\nu}$ and $J_{01\nu}$. The constants with $\mu=\lambda=\overline{2,3}$ have the same RG-flow, we further denote them as $J_{\perp\nu}$.
Without loss of generality  we take the direction of the Zeeman magnetic field along the $z$-axes. We use the approximation that the Zeeman field freezes out the spin fluctuations as soon as the running value of the field equals the high energy cutoff $\Lambda$. The condition for the running Zeeman field $h(l_h)=h_0e^{l_h}=\Lambda$ determines the logarithmic energy scale $l_h$. For the logarithmic energy scale $l<l_h$ the influence of the Zeeman field is neglected, all the spin components being equivalent.  For $l>l_h$ the structure of the RG-equation changes abruptly, the couplings with the spin-index $\nu=1,2$ stop to flow.

In what follows we absorb the density of states $\nu_F$ into the definition of coupling constants $\pi\nu_F J_{\mu\lambda\nu}\mapsto J_{\mu\lambda\nu}$. For $l<l_h$, the RG-equations for the coupling constants can be written as
\begin{eqnarray}
&&
\frac{d}{dl}Q=Q^2+R^2+\bar{J}^2, \label{Q} \\
&&
\frac{d}{dl}R=2QR, \label{R} \\
&&
\frac{d}{dl}\bar{J}=\frac{3}{2}Q\bar{J}+K\bar{J}, \label{barJ} \\
&&
\frac{d}{dl}K=\bar{J}^2. \label{K}
\end{eqnarray}
Here the following combinations of coupling constants are introduced: $Q=2(b_0J_{11\nu}+b_1J_{10\nu}$),
$R=2(b_1J_{11\nu}+b_0J_{10\nu})$, $\bar{J}=2\sqrt{b_0^2-b_1^2}J_{\perp\nu}$, and $K=b_0J_{110}+b_1J_{100}$
with $\nu=\in\{0,..., 3\}$.
Numerical solution of the RG-equations (\ref{Q})--(\ref{K}) allows to determine approximately the Kondo temperature as a function of $b_1$. The solution is shown in the inset in Fig. \ref{figTk-b1}.
One can see that without the external Zeeman field, or for Zeeman energy less than the Kondo temperature
$T_K^{SU(4)}$, the Kondo temperature diminishes only slightly between the SU(4)-symmetric  point $b_1=0$ and the SU(2) symmetric point $b_1=0.5$.   Therefore, the transition between the $SU(4)$ and $SU(2)$ Kondo effect can hardly be seen in the change of the Kondo temperature without the external Zeeman field \cite{Choi}.

\begin{figure}
\includegraphics[width=8cm,height=6cm,angle=0]{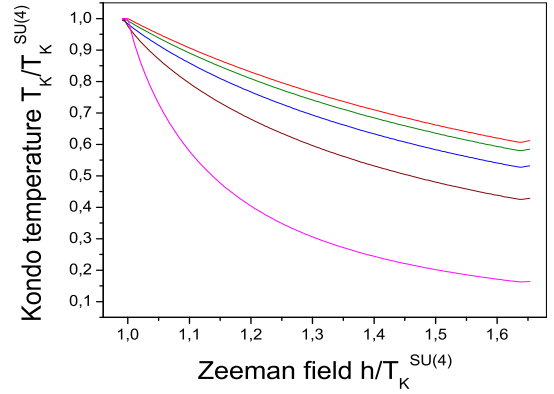}
\caption{Kondo temperature versus Zeeman magnetic field at different values of mixing $b_1$. 
From the highest to the lowest curve $b_1=0.1; 0.2; 0.3; 0.4; 0.49$.   \label{figTk-h}}
\end{figure}

For $l>l_h$ the RG equations have a form
\begin{eqnarray}
&&
\frac{d}{dl}{\mathcal K}={\mathcal J}^2, \label{Kh} \\
&&
\frac{d}{dl}{\mathcal J}={\mathcal K}{\mathcal J} \label{Jh}
\end{eqnarray}
with  ${\mathcal J}=\sqrt{b_0^2-b_1^2}(J_{\mu\mu 0}+J_{\mu\mu 3})$, $\mu=\overline{2,3}$, and
${\mathcal K}=b_0(J_{110}+J_{113})+b_1(J_{100}+J_{103})$. Eqs. (\ref{Kh}), (\ref{Jh}) are to be solved with the initial conditions given by the solution of (\ref{Q}) -- (\ref{K}) at $l=l_h$:
${\mathcal J}(0)=\bar{J}(l_h)$, ${\mathcal K}(0)=K(l_h)+Q(l_h)/2$ \cite{Martinek}. Analytical solution of (\ref{Kh}), (\ref{Jh}) results in the following dependence of the Kondo temperature on the Zeeman field
\begin{equation}
T_K(h)= h\left[\frac{{\mathcal J(0)}}{{\mathcal K}(0)+\sqrt{{\mathcal K}^2(0)-{\mathcal J}^2(0)}}
\right]^{\frac{1}{\sqrt{{\mathcal K}^2(0)-{\mathcal J}^2(0)}}}.
\label{T(h)}
\end{equation}
The dependence of the Kondo temperature on Zeeman magnetic field $h$ for different values of the mixing parameter $b_1$ and the dependence of the Kondo temperature on the parameter $b_1$ at different values of Zeeman magnetic field $h$ are shown in Figs. \ref{figTk-b1}, \ref{figTk-h}  respectively. The suppression of the Kondo temperature with Zeeman field and with the mixing $b_1$ illustrates our findings (\ref{T(h)}).

In conclusion, we considered the transition between the SU(4) and SU(2) Kondo effects in a double dot system embedded in a parallel circuit with attached leads. The transition is induced by the transfer of electrons between that dots through the leads, which violates the conservation of the pseudospin. Despite having the same Kondo temperature, the SU(4) and spin SU(2) Kondo effects can be distinguished by the sensitivity to the in plane Zeeman magnetic field. We derived the dependence of the Kondo temperature on magnetic field (\ref{T(h)}) that allows direct comparison with experiments. Being the basic energy scale, the Kondo temperature determines all other experimentally observable properties related to the Kondo effect. Therefore, the calculated dependence of the Kondo temperature of Zeeman magnetic field and the mixing parameter can be seen in the measurements of the conductance through the double dot device. The obtained results can also be relevant to the experiments on Kondo effect in carbon nanotubes \cite{Choi}.

\begin{acknowledgments}
The author thanks D. Pfannkuche, M. Trushin and F. Hellmuth for valuable comments. Support from DFG through SFB 508 is gratefully acknowledged.
\end{acknowledgments}

\subsection{}
\subsubsection{}

\end{document}